\documentclass[%
preprint,
showpacs,
 amsmath,amssymb,
 aps,
]{revtex4-1}

\usepackage{graphicx}
\usepackage{dcolumn}
\usepackage{bm}

\begin{document}

\title{Effect of Landau damping on alternative ion-acoustic solitary waves in a magnetized
plasma consisting of warm adiabatic ions and non-thermal electrons}
\author{Jayasree Das}
\affiliation{Department of Mathematics, Chittaranjan College, 8A
Beniatola Lane, Kolkata - 700 009, India.}
\author{Anup Bandyopadhyay}
\thanks{abandyopadhyay1965@gmail.com}%
\affiliation{Department of Mathematics, Jadavpur University,
Kolkata - 700 032, India.}
\author{K. P. Das}
\affiliation{Department of Applied Mathematics, University of Calcutta, 92-Acharya Profulla Chandra Road, Kolkata - 700 009,India.}

\begin{abstract}
Bandyopadhyay and Das [Phys. Plasmas, \textbf{9}, 465-473, 2002] have derived a nonlinear macroscopic evolution equation for ion acoustic wave in a magnetized plasma consisting of warm adiabatic ions and non-thermal electrons including the effect of Landau damping. In that paper they have also derived the corresponding nonlinear evolution equation when coefficient of the nonlinear term of the above mentioned macroscopic evolution equation vanishes, the nonlinear behaviour of the ion acoustic wave is described by a modified macroscopic evolution equation. But they have not considered the case when the coefficient is very near to zero. This is the case we consider in this paper and we derive the corresponding evolution equation including the effect of Landau damping. Finally, a solitary wave solution of this macroscopic evolution is obtained, whose amplitude is found to decay slowly with time.
\end{abstract}
\maketitle 

\section{\label{sec:intro}Introduction}

\begin{figure}
  \begin{center}
  \includegraphics{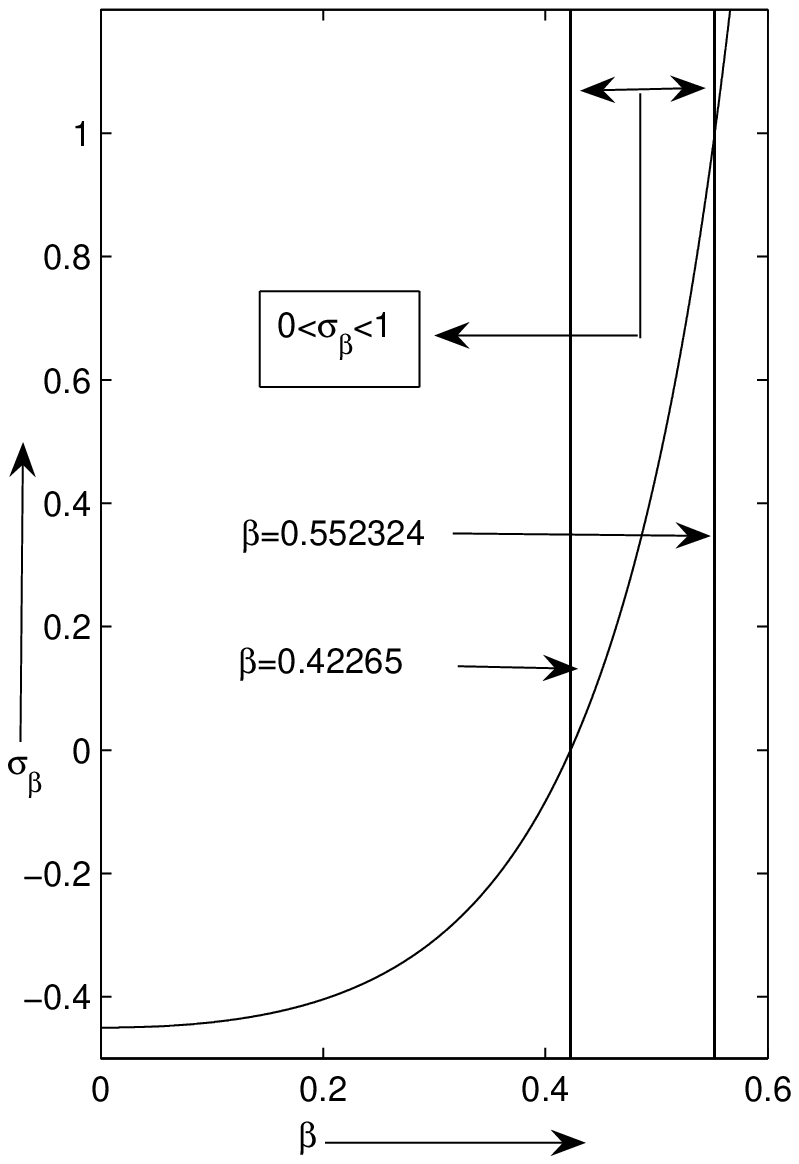}
  \caption{Variation of $\sigma_{\beta}$ against $\beta$}\label{fig1}
  \end{center}
\end{figure}

Bandyopadhyay and Das \cite{Ban02a} derived a macroscopic evolution equation to investigate the nonlinear behaviour of the ion acoustic waves in a magnetized plasma consisting of warm adiabatic ions and non-thermal electrons including the effect of Landau damping. This equation is a Korteweg-de Vries-Zakharov-Kuznetsov (KdV-ZK) equation except for an extra term that accounts for the effect of Landau damping. Bandyopadhyay and Das \cite{Ban02a} reported that this macroscopic evolution equation admits solitary wave solution propagating obliquely to the external uniform static magnetic field and having a $sech^{2}$-profile. But the amplitude of solitary wave does not remain constant; it varies slowly with time $\tau$ as $(1+\tau / \tau')^{-2}$, where $\tau'$ is a constant depending on the initial amplitude of the solitary wave, the angle between the direction of propagation of the solitary wave and the external uniform static magnetic field and the parameters involved in the system. This evolution equation, which we are discussing about, losses its validity when the coefficient of the nonlinear term of the macroscopic evolution equation vanishes and this vanishes along a particular curve in the $\beta \sigma $-parametric plane as shown in Fig.\ref{fig1}, where $\beta$ is the nonthermal parameter associated with the nonthermal distribution of electrons and $\sigma$ is the ratio of the average temperature of ions to that of electrons.  In this situation, in the same paper, they have derived a modified macroscopic evolution equation when the coefficient of the nonlinear term of the macroscopic evolution equation vanishes. This equation is a modified Korteweg-de Vries-Zakharov-Kuznetsov (MKdV-ZK) equation except for an extra term that accounts for the effect of Landau damping.  Bandyopadhyay and Das \cite{Ban02a} reported that this modified macroscopic evolution equation admits solitary wave solution propagating obliquely to the external uniform static magnetic field and having a $sech$-profile. But the amplitude of solitary wave does not remain constant; it varies slowly with time $\tau$ as $(1+\tau / \tau')^{-1}$, where $\tau'$ is a constant depending on the initial amplitude of the solitary wave, the angle between the direction of propagation of the solitary wave and the external uniform static magnetic field and the parameters involved in the system. But again this modified macroscopic evolution equation is unable to describe the nonlinear behaviour of the ion acoustic waves including the effect of Landau damping if the coefficient of the nonlinear term of the macroscopic evolution equation approaches to zero, but not exactly equal to zero. In such situation, i.e., when the coefficient of the nonlinear term of the macroscopic evolution equation approaches to zero, but not exactly equal to zero, Das \textit{et al.} \cite{Das07} derived a combined MKdV-KdV-ZK equation to describe the nonlinear behaviour of the ion acoustic wave when the Landau damping effect has not been taken into account.  Following Ott and Sudan \cite{Ott69}, in the present paper, we derive a macroscopic evolution equation to study the nonlinear behaviour of the ion acoustic waves including the effect of Landau damping. This equation is a combined MKdV-KdV-ZK equation except for an extra term that accounts for the effect of Landau damping. The solitary wave solution of this further modified macroscopic evolution has been obtained. It is found that due to inclusion of the effect of Landau damping the amplitude of the alternative solitary wave solution of this equation is a slowly varying function of time.

In the investigations made by Das \textit{et al.} \cite{Das07}, the Landau damping effect has not been taken into account. In the present paper, we include this effect on the problem considered in Das \textit{et al.} \cite{Das07}. Starting from the same governing equations but replacing the expression for the number density of non-thermal electrons by the Vlasov-Boltzmann equation for electrons, an appropriate macroscopic evolution equation corresponding to the combined modified Korteweg-de Vries-Zakharov-Kuznetsov (combined MKdV-KdV-ZK) equation of Das \textit{et al.} \cite{Das07} is derived, which describes the long-time evolution of weakly nonlinear long wave-length ion acoustic waves in a magnetized plasma consisting of warm adiabatic ions and non-thermal electrons including the effect of Landau damping.

The physics of nonlinear Landau damping is of interest for two major reasons. First, it is a fundamental and distinctive plasma phenomenon that links collective and single-particle behaviour. Second, the derivation of reduced fluid models that incorporate accurately such kinetic effects, is of great importance for plasma transport studies. For instance, some authors have proposed a $k$-dependent dissipation term, which correctly reproduces linear Landau damping within the framework of fluid models \cite{Ham90}. However, the long time behaviour of Landau damping is intrinsically nonlinear, and, in order to assess the validity of the above models, it is important to understand whether the damping will continue indefinitely, or will eventually be stopped by the nonlinearity.

The research works on solitary waves in plasmas have been done under various physical conditions such as plasmas including multi-species ions \cite{Mck04}, negative ions \cite{Gil03}, and dust particles \cite{Xue04}. In many cases, the Korteweg-de Vries (KdV) equation is used to describe basic characters of the wave. Detailed properties of the solitary waves observed in experiments in plasmas are, however, slightly different from those predicted by the
equation. Using Q-machine plasmas, Karpman \textit{et al.}  \cite{Kar80} have observed oscillations in the tail of solitary waves, which are caused by resonant particles and have shown that the tail changes its shape depending on the strength of Landau damping.

This paper is an extension of the work of Bandyopadhyay and Das \cite{Ban02a}, where we derive a further
modified macroscopic evolution equation which describe the non-linear behaviour of ion-acoustic waves in fully ionized collisionless plasma consisting of warm adiabatic ions and non-thermal electrons having vortex-like velocity distribution, immersed in a uniform static magnetic field directed along z-axis including the effect of
Landau damping. This equation is true only for the case when the coefficient of the nonlinear term of the macroscopic evolution equation derived by Bandyopadhyay and Das \cite{Ban02a} approaches to zero and not exactly equal to zero. With the help of multiple time scale analysis of Ott and Sudan \cite{Ott69}, we find a solitary wave solution of this equation. From the solution, we can conclude that the amplitude of the solitary wave slowly decreases with time.

This paper is organized as follows. The basic equation have been given in Section \ref{basic eqn}. The macroscopic evolution equations are given in Section \ref{evo eqn}, in which the derivation of MKdV-KdV-ZK like macroscopic evolution equation is given in subsection \ref{fmmee}. Solitary wave solutions of the combined MKdV-KdV-ZK like macroscopic evolution equation are investigated in Section \ref{sws}. Finally, we have concluded our findings in Section \ref{conclusion}.

\section{\label{basic eqn}Basic Equation}

The following are the governing equations describing the non-linear
behaviour of ion-acoustic waves in fully ionized collisionless
plasma consisting of warm adiabatic ions and non-thermal electrons
having vortex-like velocity distribution, immersed in a uniform
static magnetic field directed along z-axis. Here it is assumed that
the plasma beta i.e., the ratio of particle pressure to the magnetic
pressure is very small and the characteristic frequency is much
smaller than ion cyclotron frequency (Cairns \textit{et al.} \cite{Cai95c}, Mamun \cite{Mam96}).

\begin{eqnarray}\label{5.2.1}
\frac{\partial n}{\partial t}+ \nabla . (n \mathbf{u}) = 0,
\end{eqnarray}
\begin{eqnarray}\label{5.2.2}
\frac{\partial \mathbf{u}}{\partial t}+ (\mathbf{u}.\nabla)
\mathbf{u} = -\nabla \varphi + \omega_{c} ( \mathbf{u} \times
\hat{z}) - \frac{\sigma}{n} \nabla p,
\end{eqnarray}
\begin{eqnarray}\label{5.2.3}
\nabla ^{2} \varphi = n_{e} - n,
\end{eqnarray}
\begin{eqnarray}\label{5.2.4}
p = n^{\gamma}.
\end{eqnarray}
where
\begin{eqnarray}\label{5.2.5}
n_{e} = \int^{\infty} _{-\infty} f dv_{\shortparallel},
\end{eqnarray}
and the velocity distribution function of electrons $f$ must
satisfy the Vlasov- Boltzmann equation
\begin{eqnarray}\label{5.2.6}
\sqrt{\frac{m_{e}}{m_{i}}}\frac{\partial f}{\partial
t}+v_{\shortparallel} \frac{\partial f}{\partial z}+\frac{\partial
\varphi}{\partial z}\frac{\partial f}{\partial
v_{\shortparallel}}=0.\end{eqnarray}
Here $n, n_{e}, \mathbf{u}, p, \varphi, (x, y, z) $ and $t $ are
respectively the ion number density, electron number density, ion
fluid velocity, ion pressure, electrostatic potential, spatial
variables and time, and they have been normalized respectively by $n_{0}$
(unperturbed ion number density), $n_{0}$,
$c_{s}\Big(=\sqrt{\frac{K_{B}T_{e}}{m}}~\Big)$ (ion-acoustic speed), $
n_{0}K_{B}T_{i}, \frac{K_{B}T_{e}}{e},
\lambda_{D}\Big(=\sqrt{\frac{K_{B}T_{e}}{4\pi n_{0} e^{2}}}~\Big)$ (Debye
length) and $\omega_{p}^{-1}$(ion plasma period), where $\sigma =
\frac{T_{i}}{T_{e}}, \omega_{c}$ is the ion cyclotron frequency
normalized by $\omega_{p}\Big(=\sqrt{\frac{4\pi n_{0} e^{2}}{m}}~\Big)$ and
$\gamma \big(=\frac{5}{3}\big)$ is the ratio of two specific heats. Here $K_{B}$
is the Boltzmann constant; $T_{e}, T_{i}$ are respectively the
electron and ion temperatures; $m$ is the mass of an ion and $e$ is
the electronic charge $v_{\shortparallel}$ is the velocity of
electrons in phase space normalized to
$v_{e}=\sqrt{\frac{K_{B}T_{e}}{m_{e}}}$. In (\ref{5.2.4}), the
adiabatic law has been taken on the basis of the assumption that the
effect of viscosity, thermal conductivity and the energy transfer
due to collision can be neglected.

Since the electrons are assumed to be nonthermally distributed, the
electron velocity distribution function can be taken as (Cairns \textit{et al.} \cite{Cai95c})
\begin{eqnarray}\label{5.2.7}
f_{0}(v_{\shortparallel})=\frac{1}{\sqrt{2\pi}(1+3\alpha_{1})}(1+\alpha_{1}
v_{\shortparallel}^{4})e^{(-\frac{1}{2}v_{\shortparallel}^{2})}
\end{eqnarray}

To discuss the effect of Landau damping on ion-acoustic solitary
waves, we follow the method of Ott and Sudan
\cite{Ott69} and following them, we replace $\sqrt{\frac{m_{e}}{m_{i}}}$ by
$\sqrt{\frac{m_{e}}{m_{i}}}~\varepsilon$ where $\varepsilon$ is a
small parameter. The equation (\ref{5.2.6}) then assumes the following
form
\begin{eqnarray}\label{5.2.8}
\alpha_{2}\varepsilon\frac{\partial f}{\partial t}+v_{\shortparallel}
\frac{\partial f}{\partial Z}+\frac{\partial \varphi}{\partial
Z}\frac{\partial f}{\partial v_{\shortparallel}}=0,
\end{eqnarray}
where $\alpha_{2}=\sqrt{\frac{m_{e}}{m_{i}}}$.

Again using Eq. (\ref{5.2.4}), the equation (\ref{5.2.2}) becomes
\begin{eqnarray}\label{5.2.9}
\frac{\partial \mathbf{u}}{\partial t}+ (\mathbf{u}.\nabla)
\mathbf{u} = -\nabla \varphi + \omega_{c} ( \mathbf{u} \times
\hat{z}) - \sigma n^{\frac{1}{3}} \nabla n,
\end{eqnarray}
and the equation (\ref{5.2.3}), which is the Poisson equation becomes

\begin{eqnarray}\label{5.2.10}
\varepsilon\nabla_{\xi}^{2}\varphi + n-\int_{-\infty}^{\infty}fdv_{\shortparallel}=0,
\end{eqnarray}

Therefore Eqs. (\ref{5.2.1}), (\ref{5.2.9}), (\ref{5.2.10}) and (\ref{5.2.8}) are our governing equations.

\section{\label{evo eqn}Evolution equations }

\subsection{\label{mee} Macroscopic evolution equation}

Before deriving the nonlinear evolution equation for ion-acoustic
wave in a magnetized collisionless plasma consisting of warm adiabatic ions
and non-thermal electrons including the effect of Landau damping for
a particular case not considered in the paper of Bandyopadhyay and
Das \cite{Ban02a}, we give below in short a summary of the results
obtained in that paper. The macroscopic evolution equation obtained is the
following:
\begin{eqnarray}\label{5.3.8}
&& \frac{\partial \varphi^{(1)}}{\partial \tau} + A B'
\varphi^{(1)}\frac{
\partial \varphi^{(1)}}{\partial \zeta} + \frac{1}{2} A \frac{\partial ^{3}
\varphi^{(1)}}{\partial \zeta ^{3}}+ \frac{1}{2} AD
\frac{\partial}{
\partial\zeta}\Bigg(\frac{\partial ^{2} \varphi^{(1)}}{\partial \xi ^{2}} + \frac{
\partial ^{2} \varphi^{(1)}}{\partial \eta ^{2}}\Bigg)\nonumber \\
&&+\frac{1}{2}AE\alpha_{2}P \int^{\infty}_{-\infty}\frac{\partial
\varphi^{(1)}}{\partial\zeta'}\frac{d\zeta'} {\zeta-\zeta'} = 0,
\end{eqnarray}
where
\begin{eqnarray}\label{5.3.10}
A=\frac{1}{V}\Bigg(V^{2}-\frac{5 \sigma}{3}\Bigg)^{2},
\end{eqnarray}
\begin{eqnarray}\label{5.3.11}
B'=\frac{1}{2} \Bigg[
\frac{\Bigg(3V^{2}-\frac{5}{9} \sigma \Bigg)}{\Bigg(V^{2}-\frac{5}{3}\sigma \Bigg)^{3}} -1 \Bigg],
\end{eqnarray}
\begin{eqnarray}\label{5.3.12}
D=1+\frac{V^4}{\omega_{c}^2}\Bigg(V^{2}-\frac{5}{3}\sigma
\Bigg)^{-2},
\end{eqnarray}
\begin{eqnarray}\label{5.3.13}
E=\frac{V}{4\sqrt{2\pi}}(4-3\beta),
\end{eqnarray}
and the constant $V$ is given by
\begin{eqnarray}\label{5.3.14}
(1 - \beta) \Bigg(V^{2}-\frac{5}{3} \sigma \Bigg) = 1.
\end{eqnarray}
The equation (\ref{5.3.8}) is a KdV-ZK equation except for an extra term \big(last term of the left hand side of (\ref{5.3.8})\big)
that accounts for the effect of Landau damping. The solitary wave solution of the equation (\ref{5.3.8}) has
been obtained in that paper of Bandyopadhyay and Das \cite{Ban02a}. They have found that the solitary wave solution of the equation (\ref{5.3.8}) has the same $sech^{2}$ - profile as in the case of KdV-ZK equation. But, here the amplitude as well as the width of the solitary wave varies slowly with time. In particular, the amplitude ($a$) of the solitary wave solution of the equation (\ref{5.3.8})
is given by the following equation.
\begin{eqnarray}\label{5.3.15a}
a=a_{0}\bigg(1+\frac{\tau}{\tau'}\bigg)^{-2},
\end{eqnarray}
where $a_{0}$ is the value of $a$ at $\tau = 0$ and $\tau'$ is given by the following equation
\begin{eqnarray}\label{5.4.19}
\tau'&=&\Bigg[\frac{1}{4}\cos
\delta\sqrt{\frac{A^{2}E^{2}\alpha_{2}^{2}B'a_{0}}{6(\cos^{2}\delta+D\sin^{2}\delta)}}P\int_{-\infty}^{\infty}\int_{-\infty}^{\infty}sech^{2}Z\frac{\partial (sech^{2}Z')}{\partial
Z'}\frac{dZ'dZ}{Z-Z'}\Bigg]^{-1}.
\end{eqnarray}

Using (\ref{5.3.14}), the expression for $B'$ can be simplified as
\begin{eqnarray}\label{5.315}
B'=\frac{20(1-\beta)^3}{9}(\sigma-\sigma_{\beta}),
\end{eqnarray}
where
\begin{eqnarray}\label{5.315a}
\sigma_{\beta}=\frac{9\{1-3(1-\beta)^2\}}{40(1-\beta)^3}.
\end{eqnarray}

From this expression of $B'$, it is easy to see that the coefficient of the non-linear term in (\ref{5.3.8}) vanishes along a particular curve Fig.\ref{fig1} in the $\beta \sigma$ plane, and consequently, it is not possible to discuss the nonlinear behaviour of ion acoustic wave including the effect of Landau damping with the help of Eq. (\ref{5.3.8}). In this situation, i.e., when $B'=0$, Bandyopadhyay and Das \cite{Ban02a} have also derived a modified macroscopic evolution equation.

\subsection{\label{mmee}Modified macroscopic evolution equation}

For this case, i.e., when $B' = 0$, giving appropriate stretching of space coordinates
and time, and appropriate perturbation expansions of the dependent variables
Bandyopadhyay and Das \cite{Ban02a} in the same paper have derived the following modified macroscopic evolution
equation for ion acoustic waves in a fully ionized collisionless
plasma consisting of warm ions and non-thermal electrons immersed in
a uniform static magnetic field directed along the z-axis:
\begin{eqnarray}\label{5.3.17}
&& \frac{\partial \varphi^{(1)}}{\partial \tau} + A B''
[\varphi^{(1)}]^2\frac{\partial \varphi^{(1)}}{\partial \zeta} +
\frac{1}{2} A \frac{\partial ^{3} \varphi^{(1)}}{\partial \zeta
^{3}} + \frac{1}{2} AD
\frac{\partial}{\partial\zeta}\Bigg(\frac{\partial ^{2}
\varphi^{(1)}}{\partial \xi ^{2}} + \frac{\partial ^{2}
\varphi^{(1)}}{\partial \eta ^{2}}\Bigg)\nonumber \\
&&+\frac{1}{2}AE\alpha_{2}P \int^{\infty}_{-\infty}\frac{\partial
\varphi^{(1)}}{\partial\zeta'}\frac{d\zeta'} {\zeta-\zeta'} = 0.
\end{eqnarray}
Here $A$, $D$ and $E$ are same as given by the equations (\ref{5.3.10}), (\ref{5.3.12})
and (\ref{5.3.13}) respectively and $B''$ is given by the following
equation:
\begin{eqnarray}\label{5.3.18}
 B'' &=& \frac{1}{2}\bigg[\bigg(V^{2}-\frac{5}{3} \sigma \bigg)^{-4}\bigg\{\frac{10}{27} \sigma -6V^{2}+\frac{3}{2}\bigg(V^{2}- \frac{5}{3} \sigma \bigg)^{-1}\bigg(3V^{2}-\frac{5}{9} \sigma \bigg)\bigg\}\nonumber \\
 &&-\frac{1}{2}(1+3\beta)\bigg],
\end{eqnarray}
and the constant $V$ is determined from equation (\ref{5.3.14}).

The equation (\ref{5.3.17}) is a MKdV-ZK equation except for an extra term \big(last term of the left hand side of  (\ref{5.3.17})\big)
that accounts for the effect of Landau damping. The solitary wave solution of the equation (\ref{5.3.17}) has
been investigated by Bandyopadhyay and Das \cite{Ban02a} in the same paper. They have found that the solitary wave solution of the
equation (\ref{5.3.17}) has the same $sech$-profile as in the case of MKdV-ZK equation. But, here the amplitude as well as the width of the solitary wave varies slowly with time. In particular, the amplitude ($a$) of the solitary wave solution of the equation (\ref{5.3.17})
is given by the following equation.
\begin{eqnarray}\label{5.3.15}
a=a_{0}\bigg(1+\frac{\tau}{\tau'}\bigg)^{-1},
\end{eqnarray}
where $a_{0}$ is the value of $a$ at $\tau = 0$ and $\tau'$ is given by the following equation
\begin{eqnarray}\label{5.4.22}
\tau'&=&\Bigg[\frac{1}{2}\cos
\delta\sqrt{\frac{A^{2}E^{2}\alpha_{2}^{2}a_{0}^{2}B''}{3(\cos^{2}\delta+D\sin^{2}\delta)}}
P\int_{-\infty}^{\infty}\int_{-\infty}^{\infty}sechZ\frac{\partial (sechZ')}{\partial
Z'}\frac{dZ'dZ}{Z-Z'}\Bigg]^{-1}.
\end{eqnarray}

But both the evolution equations (\ref{5.3.8}) and (\ref{5.3.17}) are unable to describe the nonlinear behaviour of the ion acoustic wave along with the effect of Landau damping in the neighbourhood of the curve $\sigma=\sigma_{\beta}$ in the $\beta\sigma$ - parametric plane along which $B'=0$ (Fig.\ref{fig1}). This is the situation we are considering here. We have derive in this case, a further modified macroscopic evolution equation which describes the nonlinear behaviour of the ion acoustic wave in the neighbourhood of the curve $\sigma=\sigma_{\beta}$ in the $\beta\sigma$ - parametric plane along which $B'=0$.

\subsection{\label{fmmee} Further Modified Macroscopic evolution equation}

To discuss the nonlinear behaviour of the ion acoustic wave in the neighbourhood
of the curve in the $\beta \sigma$ parametric plane along which $B'=0$, we assume $B'\approx\bigcirc(\epsilon^\frac{1}{2})$ (Nejoh \cite{Nej92}),
and we take the following stretching of space coordinates
and time.
\begin{eqnarray}\label{5.3.1}
\xi=\varepsilon^\frac{1}{2}x,\eta=\varepsilon^\frac{1}{2}y,\zeta=\varepsilon^\frac{1}{2}(z-Vt),\tau=\varepsilon^\frac{3}{2}t,~
\end{eqnarray}
where $\varepsilon$ is a small parameter measuring the weakness of
the dispersion and $V$ is a constant.

With the stretching given by (\ref{5.3.1}), the equations
(\ref{5.2.1}), (\ref{5.2.9}), (\ref{5.2.10}) and (\ref{5.2.8})
respectively assume the following form:
\begin{eqnarray}\label{5.3.2}
&&-\varepsilon^\frac{1}{2}V\frac{\partial n}{\partial\zeta}+
\varepsilon^\frac{3}{2}\frac{\partial n}{\partial
\tau}+\varepsilon^\frac{1}{2}\mathbf{\nabla}_{\xi}.(n\mathbf{u})=0,
\end{eqnarray}
\begin{eqnarray}\label{5.3.3}
&&-\varepsilon^\frac{1}{2}V\frac{\partial \mathbf{u}}{\partial
\zeta}+\varepsilon^\frac{3}{2}\frac{\partial \mathbf{u}}{\partial
\tau}+\varepsilon^\frac{1}{2}(\mathbf{u}.\mathbf{\nabla}_{\xi})\mathbf{u} =
-\varepsilon^\frac{1}{2}\mathbf{\nabla}_{\xi}\varphi +\omega_{c}(\mathbf{u}\times\widehat{z})-\frac{5}{3}\sigma\varepsilon^\frac{1}{2}
n^{-\frac{1}{3}}\mathbf{\nabla}_{\xi}n,
\end{eqnarray}
\begin{eqnarray}\label{5.3.4}
&&\varepsilon\nabla_{\xi}^{2}\varphi + n-\int_{-\infty}^{\infty}fdv_{\shortparallel}=0,
\end{eqnarray}
\begin{eqnarray}\label{5.3.6}
&&-V\alpha_{2}\varepsilon^{\frac{3}{2}}\frac{\partial
f}{\partial\zeta}+\alpha_{1}\varepsilon^{\frac{5}{2}}\frac{\partial
f}{\partial\tau}+\varepsilon^{\frac{1}{2}}v_{\shortparallel}\frac{\partial
f}{\partial\zeta}+\varepsilon^{\frac{1}{2}}\frac{\partial
\varphi}{\partial\zeta}\frac{\partial f}{\partial v_{\shortparallel}}=0.
\end{eqnarray}
Here
\begin{eqnarray}\label{5.3.7}
\mathbf{\nabla}_{\xi}=\widehat{e}_{x}\frac{\partial}{\partial
\xi}+\widehat{e}_{y}\frac{\partial}{\partial
\eta}+\widehat{e}_{z}\frac{\partial}{\partial \zeta},
\mathbf{u}=(u,v,w).~~
\end{eqnarray}
Next we use the following perturbation expansions of the dependent
variables to make a balance between nonlinear and dispersive terms.
\begin{eqnarray}\label{5.3.180}
\left.\begin{array}{ll}(n,\varphi, w, f) = (1, 1, 0, 0,
f_{0})+\sum_{i=1}^{\infty}\varepsilon^{\frac{i}{2}}(n^{(i)},
\varphi^{(i)}, w^{(i)}, f^{(i)}) ,\\\\
(u,v) = \sum_{i=1}^{\infty}\varepsilon^{\frac{i+1}{2}}(u^{(i)}, v^{(i)}).~~~~
\end{array} \right\}
\end{eqnarray}
Substituting (\ref{5.3.180}) into the equations (\ref{5.3.2})-(\ref{5.3.6}) and then
equating coefficient of different powers of $\varepsilon$ on both sides, we get a sequence of equations. From the
lowest order equations obtained from (\ref{5.3.2})-(\ref{5.3.4}), which are at the order $\epsilon$, we get the following equations.
\begin{eqnarray}\label{a10}
\left.\begin{array}{ll}n^{(1)} = \bigg(V^{2}-\frac{5}{3} \sigma \bigg)^{-1} \varphi^{(1)},\\
w^{(1)} = V \bigg(V^{2}-\frac{5}{3} \sigma \bigg)^{-1} \varphi^{(1)},\\
u^{(1)} = - \frac{V^{2}}{\omega_{c}}\bigg(V^{2}-\frac{5}{3} \sigma \bigg)^{-1} \frac{\partial\varphi^{(1)}}{\partial \eta},\\
v^{(1)} = \frac{V^{2}}{\omega_{c}}\bigg(V^{2}-\frac{5}{3} \sigma \bigg)^{-1} \frac{\partial\varphi^{(1)}}{\partial \xi},\\
n^{(1)} = \int_{-\infty}^{\infty}f^{(1)}dv_{\shortparallel}.~~~~~
\end{array} \right\}
\end{eqnarray}
From the Vlasov equation (\ref{5.3.6}) at the lowest order, i.e., at the order $\epsilon^{1/2}$
, we get the following equation
\begin{eqnarray}\label{a20}
v_{\shortparallel}\frac{\partial f^{(1)}}{\partial \zeta} + \frac{\partial f_{0}}{\partial v_{\shortparallel}}\frac{\partial \varphi^{(1)}}{\partial \zeta}=0.
\end{eqnarray}
As this equation does not have a unique solutions, we include an extra higher order term $\varepsilon^{3}\alpha_{2}\frac{\partial f^{(1)}}{\partial \tau}=\varepsilon \varepsilon^{2}\alpha_{2}\frac{\partial f^{(1)}}{\partial \tau}$ originated from the Vlasov equation at the order $\varepsilon^{2}$. Let us write the equation
(\ref{a20}) as follows.
\begin{eqnarray}\label{a30}
\varepsilon^{2}\alpha_{2}\frac{\partial f_{\varepsilon}^{(1)}}{\partial \tau}+v_{\shortparallel}\frac{\partial f_{\varepsilon}^{(1)}}{\partial \zeta} + \frac{\partial f_{0}}{\partial v_{\shortparallel}}\frac{\partial \varphi^{(1)}}{\partial \zeta}=0.
\end{eqnarray}
Then $f^{(1)}$ can be obtained as unique  solution of this equation by imposing the natural relation of the form
\begin{eqnarray}\label{a40}
f^{(1)}=\lim_{\varepsilon \rightarrow 0} f_{\varepsilon}^{(1)}.
\end{eqnarray}
Assuming $\tau$ dependence of $f_{\varepsilon}^{(1)}$ and $\varphi^{(1)}$ to be of the form $exp(i \omega \tau)$, the equation (\ref{a40}) can be written as
\begin{eqnarray}\label{a50}
i \omega \varepsilon^{2}\alpha_{2}f_{\varepsilon}^{(1)}+v_{\shortparallel}\frac{\partial f_{\varepsilon}^{(1)}}{\partial \zeta} + \frac{\partial f_{0}}{\partial v_{\shortparallel}}\frac{\partial \varphi^{(1)}}{\partial \zeta}=0.
\end{eqnarray}
Now taking Fourier transform of this equation with respect to the variable $\zeta$ according to the definition,
\begin{eqnarray}\label{a60}
\hat{g}=\hat{g}(k)=\frac{1}{\sqrt{2\pi}}\int_{-\infty}^{\infty}g(\zeta)exp(-ik\zeta)d \zeta,
\end{eqnarray}
we get
\begin{eqnarray}\label{a70}
i \omega \varepsilon^{2}\alpha_{2}\hat{f}_{\varepsilon}^{(1)}+i k v_{\shortparallel}\hat{f}_{\varepsilon}^{(1)} + i k \frac{\partial f_{0}}{\partial v_{\shortparallel}}\hat{\varphi}^{(1)}=0.
\end{eqnarray}
This equation gives the following expression for $\hat{f}_{\varepsilon}^{(1)}$:
\begin{eqnarray}\label{a80}
\hat{f}_{\varepsilon}^{(1)}= -\frac{k \frac{\partial f_{0}}{\partial v_{\shortparallel}}}{k v_{\shortparallel}+\alpha_{2}\omega \varepsilon^{2}}\hat{\varphi}^{(1)}.
\end{eqnarray}
Now whenever the factor $1/(k v_{\shortparallel}+\alpha_{2}\omega \varepsilon^{2})$ comes under an integration over $v_{\shortparallel}$ along the real axis, the general prescription is to replace this integration according to Landau, along a contour in the complex $v_{\shortparallel}$-plane known as Landau contour. This is equivalent to replacing the factor $1/(k v_{\shortparallel}+\alpha_{2}\omega \varepsilon^{2})$ by the following
\begin{eqnarray}\label{a90}
\frac{1}{k v_{\shortparallel}+\alpha_{2}\omega \varepsilon^{2}}&=&P \frac{1}{k v_{\shortparallel}+\alpha_{2}\omega \varepsilon^{2}}+i\pi \delta(k v_{\shortparallel}+\alpha_{2}\omega \varepsilon^{2}).
\end{eqnarray}
Substituting this relation into the equation (\ref{a80}) and then proceeding to the limit $\varepsilon \longrightarrow 0+$, we get according to (\ref{a40}) the following expression for $\hat{f}^{(1)}$:
\begin{eqnarray}\label{a300}
\hat{f}^{(1)}=- 2\frac{\partial f_{0}}{\partial v_{\shortparallel}^{2}} \hat{\varphi}^{(1)}-
2 \pi i \frac{\partial f_{0}}{\partial v_{\shortparallel}^{2}} k v_{\shortparallel} \delta(k v_{\shortparallel}) .
\end{eqnarray}
Due to the relation $x\delta(x)=0$, this equation assumes the following form:
\begin{eqnarray}\label{a310}
\hat{f}^{(1)}=- 2\frac{\partial f_{0}}{\partial v_{\shortparallel}^{2}} \hat{\varphi}^{(1)}.
\end{eqnarray}
Taking Fourier inverse transform of (\ref{a310}), we get
\begin{eqnarray}\label{a320}
f^{(1)}=- 2\frac{\partial f_{0}}{\partial v_{\shortparallel}^{2}} \varphi^{(1)}.
\end{eqnarray}
Substituting (\ref{a320}) in the last equation of (\ref{a10}) and then performing the integration we get
\begin{eqnarray}\label{a330}
n^{(1)} = (1-\beta) \varphi^{(1)} \mbox{  with  } \beta = \frac{4\alpha_{1}}{1+3\alpha_{1}}.
\end{eqnarray}
This equation along with the first equation of (\ref{a10}) gives the following dispersion relation to determine the constant $V$
\begin{eqnarray}\label{a340}
(1-\beta)\bigg(V^{2}-\frac{5}{3} \sigma \bigg)=1.
\end{eqnarray}
This equation is same as the equation (\ref{5.3.14}) as well as the equation (20) of Das et. al. \cite{Das07}.
In the next order, i.e., at the order $\varepsilon^{3/2}$, solving the ion continuity equation and the parallel component (i.e., the component parallel to the ambient magnetic field, i.e., z-component or $\zeta$-component) of ion fluid equation of motion for $n^{(2)}$ and $w^{(2)}$ to express them in terms of $\varphi^{(1)}$ and $\varphi^{(2)}$, we get the following equations:
\begin{eqnarray}\label{a350}
\left.\begin{array}{ll} 
n^{(2)} =\frac{1}{\bigg(V^{2}-\frac{5}{3} \sigma \bigg)} \varphi^{(2)}
+\frac{1}{2}\frac{\bigg(3V^{2}-\frac{5}{9} \sigma \bigg)}{\bigg(V^{2}-\frac{5}{3} \sigma \bigg)^{3}} [\varphi^{(1)}]^{2},\\
w^{(2)} = \frac{V}{\bigg(V^{2}-\frac{5}{3} \sigma \bigg)}\varphi^{(2)}
+\frac{1}{2}V \frac{\bigg(V^{2}+\frac{25}{9} \sigma \bigg)}{\bigg(V^{2}-\frac{5}{3} \sigma \bigg)^{3}} [\varphi^{(1)}]^{2}.\end{array}\right\}
\end{eqnarray}
From the perpendicular component (i.e., the component perpendicular to the ambient magnetic field, i.e., the components along x-axis and y-axis) of the ion fluid equation of motion at the order
$\varepsilon^{3/2}$, we get the following equation:
\begin{eqnarray}\label{a360}
\frac{\partial u^{(2)}}{\partial \xi} +\frac{\partial v^{(2)}}{\partial \eta}&=& \frac{V^{3}}{\omega_{c}^{2}}\bigg(V^{2}-\frac{5}{3} \sigma \bigg)^{-1} \frac{\partial }{\partial \zeta}\bigg(\frac{\partial^{2} \varphi^{(1)}}{\partial \xi^{2}} + \frac{\partial^{2} \varphi^{(1)}}{\partial \eta^{2}}\bigg).
\end{eqnarray}
From the Poisson equation at the order $\varepsilon$, we get
\begin{eqnarray}\label{a370}
n^{(2)} - \int_{-\infty}^{\infty}f^{(2)}dv_{\shortparallel}=0.
\end{eqnarray}
To find $f^{(2)}$, we again consider the Vlasov equation at the order $\varepsilon^{3/2}$. The Vlasov equation at the order $\varepsilon^{3/2}$ is the following, in which as mentioned in the lowest order Vlasov equation, an extra time derivative term $\varepsilon^{7/2}\alpha_{2}\frac{\partial f^{(2)}}{\partial \tau}=\varepsilon^{3/2} \varepsilon^{2}\alpha_{2}\frac{\partial f^{(2)}}{\partial \tau}$ has been included and $f^{(2)}$ has been replaced by   $f_{\varepsilon}^{(2)}$.
\begin{eqnarray}\label{a380}
&&\varepsilon^{2}\alpha_{2}\frac{\partial f_{\varepsilon}^{(2)}}{\partial \tau}+v_{\shortparallel}\frac{\partial f_{\varepsilon}^{(2)}}{\partial \zeta} + \frac{\partial f_{0}}{\partial v_{\shortparallel}}\frac{\partial \varphi^{(2)}}{\partial \zeta}=2v_{\shortparallel}\frac{\partial^{2}f_{0}}{\partial (v_{\shortparallel}^{2})^{2}}\frac{\partial}{\partial \zeta}[\varphi^{(1)}]^{2}.
\end{eqnarray}
Then $f^{(2)}$ can be obtained as unique  solution of this equation
by imposing the natural relation of the form
\begin{eqnarray}\label{a390}
f^{(2)}=\lim_{\varepsilon \rightarrow 0+} f_{\varepsilon}^{(2)}.
\end{eqnarray}
Assuming $\tau$ dependence of $f_{\varepsilon}^{(2)}$ and $\varphi^{(2)}$ to be of the form $exp(i \omega \tau)$, taking Fourier transform of this equation with respect to the variable $\zeta$, using the causality condition (\ref{a90}) and finally proceeding to the limit $\varepsilon \rightarrow 0+$, we get according to (\ref{a390}) the following expression for $\hat{f}^{(2)}$:
\begin{eqnarray}\label{a400}
\hat{f}^{(2)}=- 2\frac{\partial f_{0}}{\partial v_{\shortparallel}^{2}} \hat{\varphi}^{(2)}
+2\frac{\partial^{2}f_{0}}{\partial (v_{\shortparallel}^{2})^{2}} \hat{d}_{3},
\end{eqnarray}
where
\begin{eqnarray}\label{a410}
\hat{d}_{3}=[ \hat{\varphi}^{(1)}]^{2},
\end{eqnarray}
and we have used the relations $x\delta(x)=0$ and $xP(\frac{1}{x})=1$ to simplify the equation (\ref{a400}).

Taking Fourier inverse transform of (\ref{a400}), we get
\begin{eqnarray}\label{a420}
f^{(2)}=- 2\frac{\partial f_{0}}{\partial v_{\shortparallel}^{2}} \varphi^{(2)}+2\frac{\partial^{2}f_{0}}{\partial (v_{\shortparallel}^{2})^{2}}[ \varphi^{(1)}]^{2}.
\end{eqnarray}
Substituting for $f^{(2)}$ and $n^{(2)}$ given respectively by (\ref{a420}) and the first equation of (\ref{a350}) into the equation (\ref{a370}), we get the following equation after simplification
\begin{eqnarray}\label{a430}
&&-\bigg(V^{2}-\frac{5}{3} \sigma \bigg)^{-1}\bigg[(1-\beta)\bigg(V^{2}-\frac{5}{3} \sigma \bigg)-1 \bigg]\varphi^{(2)}
+B' [\varphi^{(1)}]^{2}=0.
\end{eqnarray}
Now the first term of the left hand side of the equation (\ref{a430}) is identically equal to zero due to the  dispersion relation as given by the equation (\ref{a340}) and as $B'\approx$ O($\varepsilon^{1/2}$), the second term of the left hand side of the equation (\ref{a430}) along with its sign has to be included in the next higher order Poisson equation, i.e., this term along with its sign must be included in the left hand side Poisson equation of order $\varepsilon^{3/2}$ and consequently the equation (\ref{a430}), i.e., the Poisson equation of order $\varepsilon^{2/2}(=\varepsilon)$ is identically satisfied. So, including the term $+B'[\varphi^{(1)}]^{2}$ in the Poisson equation of order $\varepsilon^{3/2}$, we can write the Poisson equation at the order $\varepsilon^{3/2}$ as follows
\begin{eqnarray}\label{a440}
&&\frac{\partial^{2}\varphi^{(1)}}{\partial \xi^{2}}+\frac{\partial^{2}\varphi^{(1)}}{\partial \eta^{2}}+\frac{\partial^{2}\varphi^{(1)}}{\partial \zeta^{2}}+n^{(3)} - \int_{-\infty}^{\infty}f^{(3)}dv_{\shortparallel}+B'[\varphi^{(1)}]^{2}=0.
\end{eqnarray}

Now, in the next order, i.e., at the order $\varepsilon^{2}$, solving the ion continuity equation and the parallel component of ion fluid equation of motion for the variable $\frac{\partial n^{(3)}}{\partial \zeta}$ to express it in terms of $\varphi^{(3)}$, $\varphi^{(2)}$, $\varphi^{(1)}$, we get the following equation.
\begin{eqnarray}\label{a450}
\frac{\partial n^{(3)}}{\partial \zeta} &=& \bigg(V^{2}-\frac{5}{3} \sigma \bigg)^{-1}\frac{\partial \varphi^{(3)}}{\partial \zeta}
+  2V\bigg(V^{2}-\frac{5}{3} \sigma \bigg)^{-2}\frac{\partial \varphi^{(1)}}{\partial \tau}\nonumber \\
&+& \frac{V^{4}}{\omega_{c}^{2}\bigg(V^{2}-\frac{5}{3} \sigma \bigg)^{2}}\frac{\partial}{\partial \zeta}\bigg[\frac{\partial^{2}\varphi^{(1)}}{\partial \xi^{2}}+
\frac{\partial^{2}\varphi^{(1)}}{\partial \eta^{2}}\bigg]\nonumber \\
&+& \bigg(V^{2}-\frac{5}{3} \sigma \bigg)^{-4}\Bigg[\frac{10}{27} \sigma -6V^{2}+ \frac{3}{2}\frac{\bigg(3V^{2}-\frac{5}{9} \sigma \bigg)}{\bigg(V^{2}-\frac{5}{3} \sigma \bigg)}\Bigg][\varphi^{(1)}]^{2}\frac{\partial \varphi^{(1)}}{\partial \zeta}\nonumber \\
&+& \frac{\bigg(3V^{2}-\frac{5}{9} \sigma \bigg)}{\bigg(V^{2}-\frac{5}{3} \sigma \bigg)^{3}}\frac{\partial }{\partial \zeta}[\varphi^{(1)}\varphi^{(2)}],
\end{eqnarray}
where we have used equations (\ref{a10}), (\ref{a350}) and (\ref{a360}) to get this equation in this present form.

Now our task is to find $f^{(3)}$ that determines $n^{(3)}$ from the Poisson equation (\ref{a440}) at the order $\varepsilon^{3/2}$. To find $f^{(3)}$ we consider the Vlasov equation at the order $\varepsilon^{2}$.  The Vlasov equation at the order $\varepsilon^{2}$ is the following, in which as in the lowest order case an extra higher order term $\varepsilon^{4} \alpha_{2} \frac{\partial f^{(3)}}{\partial \tau}$ has been included and $f^{(3)}$ has been replaced by $f_{\varepsilon}^{(3)}$ and where we have substituted the expressions for $f^{(1)}$ and $f^{(2)}$ given by equations (\ref{a320}) and (\ref{a420}) respectively.
\begin{eqnarray}\label{a460}
\varepsilon^{2}\alpha_{2}\frac{\partial f_{\varepsilon}^{(3)}}{\partial \tau}+v_{\shortparallel}\frac{\partial f_{\varepsilon}^{(3)}}{\partial \zeta} + \frac{\partial f_{0}}{\partial v_{\shortparallel}}\frac{\partial \varphi^{(3)}}{\partial \zeta} = -2V \alpha_{2}\frac{\partial f_{0}}{\partial v_{\shortparallel}^{2}}x_{2}
+ 4v_{\shortparallel}\frac{\partial^{2} f_{0}}{\partial (v_{\shortparallel}^{2})^{2}}y_{2}-4v_{\shortparallel}\frac{\partial^{3} f_{0}}{\partial (v_{\shortparallel}^{2})^{3}}z_{2},
\end{eqnarray}
where
\begin{eqnarray}\label{a470}
\left.\begin{array}{lllll} 
x_{2} = \frac{\partial \varphi^{(1)}}{\partial \zeta}, \\\\
y_{2} = \frac{\partial }{\partial \zeta}[\varphi^{(1)}\varphi^{(2)}], \\\\
z_{2} = [\varphi^{(1)}]^{2}\frac{\partial \varphi^{(1)}}{\partial \zeta}.
\end{array}\right\}
\end{eqnarray}
Therefore $f^{(3)}$ is obtained from the unique solution of the equation (\ref{a460}) by the relation
\begin{eqnarray}\label{a480}
f^{(3)} = \lim_{\varepsilon \rightarrow 0+}f_{\varepsilon}^{(3)}.
\end{eqnarray}
As in the earlier cases, assuming $\tau$ dependence of
$f_{\varepsilon}^{(3)}$ and $\varphi^{(3)}$ to be of the form $exp(i
\omega \tau)$, taking Fourier transform of this equation with
respect to the variable $\zeta$, using the causality condition
(\ref{a90}) and finally proceeding to the limit $\varepsilon
\rightarrow 0+$, we get according to (\ref{a390}) the following
equation determining $\hat{f}^{(3)}$:
\begin{eqnarray}\label{a490}
ik \bigg [\hat{f}^{(3)}+2\frac{\partial f_{0}}{\partial v_{\shortparallel}^{2}} \hat{\varphi}^{(3)}\bigg] &=& - 2V \alpha_{2}\frac{\partial f_{0}}{\partial v_{\shortparallel}^{2}}\bigg[kP\bigg(\frac{1}{kv_{\shortparallel}}\bigg)+i\pi sgn(k)\delta(v_{\shortparallel})\bigg]\hat{x}_{2}\nonumber \\
&&+ 4\frac{\partial^{2}f_{0}}{\partial (v_{\shortparallel}^{2})^{2}} \hat{y}_{2}
-4\frac{\partial^{3}f_{0}}{\partial (v_{\shortparallel}^{2})^{3}} \hat{z}_{2}.
\end{eqnarray}
Integrating (\ref{a490}) over the entire range of $v_{\shortparallel}$, we get the following equation.
\begin{eqnarray}\label{a500}
ik [\hat{n}_{e}^{(3)}-(1-\beta) \hat{\varphi}^{(3)}] &=& - \frac{1}{4}iV \alpha_{2}(4-3\beta)\sqrt{\frac{\pi}{2}}sgn(k)\hat{x}_{2}+ \hat{y}_{2}+\frac{1}{2}(1+3\beta) \hat{z}_{2},
\end{eqnarray}
where we set
\begin{eqnarray}\label{a510}
n_{e}^{(3)}=\int_{-\infty}^{\infty}f^{(3)}dv_{\shortparallel}.
\end{eqnarray}
Taking inverse Fourier transform of the above equation, we get
\begin{eqnarray}\label{a520}
\frac{\partial n_{e}^{(3)}}{\partial \zeta} &=& (1-\beta) \frac{\partial \varphi^{(3)}}{\partial \zeta}
+\frac{\partial }{\partial \zeta}(\varphi^{(1)}\varphi^{(2)})+\frac{1}{2}(1+3\beta) [\varphi^{(1)}]^{2} \frac{\partial \varphi^{(1)}}{\partial \zeta}\nonumber \\
&&- \frac{V}{4 \sqrt{2 \pi}} \alpha_{2}(4-3\beta)P \int_{-\infty}^{\infty} \frac{\partial \varphi^{(1)}}{\partial \zeta'} \frac{d \zeta'}{\zeta-\zeta'},
\end{eqnarray}
in which the convolution theorem has been used to find the inverse Fourier transform of $sgn(k)\hat{x_{2}}$. Now using the equations (\ref{a510}) and (\ref{a520}), we get the following equation.
\begin{eqnarray}\label{a530}
\frac{\partial }{\partial \zeta}\bigg[\int_{-\infty}^{\infty}f^{(3)}dv_{\shortparallel}\bigg] &=& (1-\beta) \frac{\partial \varphi^{(3)}}{\partial \zeta}
+\frac{\partial }{\partial \zeta}(\varphi^{(1)}\varphi^{(2)})
+\frac{1}{2}(1+3\beta) [\varphi^{(1)}]^{2} \frac{\partial \varphi^{(1)}}{\partial \zeta}\nonumber \\
&&- \frac{1}{4 \sqrt{2 \pi}}V \alpha_{2}(4-3\beta)P \int_{-\infty}^{\infty} \frac{\partial \varphi^{(1)}}{\partial \zeta'} \frac{d \zeta'}{\zeta-\zeta'},
\end{eqnarray}
Substituting (\ref{a530}) into the equation obtained by differentiating the Poisson equation (\ref{a440}) at the order $\varepsilon^{3/2}$ with respect to $\zeta$, we get the following equation
\begin{eqnarray}\label{a540}
&& \frac{\partial n^{(3)}}{\partial \zeta}-(1-\beta)\frac{\partial \varphi^{(3)}}{\partial \zeta}+2B'\varphi^{(1)}\frac{\partial \varphi^{(1)}}{\partial \zeta}-\frac{1}{2}(1+3\beta) [\varphi^{(1)}]^{2} \frac{\partial \varphi^{(1)}}{\partial \zeta}\nonumber \\
&& +\frac{\partial}{\partial \zeta}\bigg[\frac{\partial^{2}\varphi^{(1)}}{\partial \xi^{2}}+\frac{\partial^{2}\varphi^{(1)}}{\partial \eta^{2}}+\frac{\partial^{2}\varphi^{(1)}}{\partial \zeta^{2}}\bigg] +\frac{1}{4 \sqrt{2 \pi}}V \alpha_{2}(4-3\beta)P \int_{-\infty}^{\infty} \frac{\partial \varphi^{(1)}}{\partial \zeta'} \frac{d \zeta'}{\zeta-\zeta'}\nonumber\\&&-\frac{\partial }{\partial \zeta}(\varphi^{(1)}\varphi^{(2)})=0.
\end{eqnarray}

Now substituting for $\frac{\partial n^{(3)}}{\partial \zeta}$ given by (\ref{a450}) into the equation (\ref{a540}), we get the following further modified macroscopic evolution equation, where the term $+2B'\frac{\partial}{\partial \zeta}(\varphi^{(1)}\varphi^{(2)})$ being of higher order since $B'=\bigcirc(\epsilon^{1/2})$ has been omitted.
\begin{eqnarray}\label{5.3.19}
&&\frac{\partial \varphi^{(1)}}{\partial \tau} + AB'
\varphi^{(1)}\frac{\partial \varphi^{(1)}}{\partial \zeta} + AB''
[\varphi^{(1)}]^2\frac{
\partial \varphi^{(1)}}{\partial \zeta} +\frac{1}{2} A \frac{\partial ^{3} \varphi^{(1)}}{\partial \zeta
^{3}}\nonumber\\&&+ \frac{1}{2} AD \frac{\partial}{\partial\zeta}\Bigg(\frac{\partial
^{2} \varphi^{(1)}}{\partial \xi ^{2}} + \frac{\partial ^{2}
\varphi^{(1)}}{\partial \eta ^{2}}\Bigg)+\frac{1}{2}AE\alpha_{2}P
\int^{\infty}_{-\infty}\frac{\partial
\varphi^{(1)}}{\partial\zeta'}\frac{d\zeta'} {\zeta-\zeta'} =0.
\end{eqnarray}
Here $A, B', D$ and $B''$ are respectively given by the equations (\ref{5.3.10})-(\ref{5.3.13}) and (\ref{5.3.18}) and the constant $V$ is determined by the equation (\ref{5.3.14}). The equation (\ref{5.3.19}) is a combined MKdV-KdV-ZK equation except for an extra term \big(last term of the left hand side of (\ref{5.3.19})\big) that accounts for the effect of Landau damping. In the next section, we find the solitary wave solution of this further modified macroscopic evolution equation.

\section{\label{sws}Solitary wave solution of the further modified macroscopic equation}

If we neglect the electron to ion mass ratio, i.e., if we set
$\alpha_{2}=0$, the equation (\ref{5.3.19}) reduce to a
combined MKdV-KdV-ZK equation. The solitary wave solution of this combined MKdV-KdV-ZK equation has
been studied in Das \textit{et al.} \cite{Das07}. In this paper, our aim is to find the solitary wave solution of the equation (\ref{5.3.19}).

The solitary wave solution of the equation (\ref{5.3.19}) with $\alpha_{2}=0$ propagating at an angle $\delta$ with
the external uniform static magnetic field is the following, which has already been obtained in section IV of Das et al. \cite{Das07} ,
\begin{eqnarray}\label{5.4.6}
\varphi ^{(1)} =\varphi _{0}(Z)=a\frac{S}{\Psi}.
\end{eqnarray}
where
\begin{eqnarray}\label{5.4.8}
S=sech[2pZ],
\end{eqnarray}
\begin{eqnarray}\label{5.4.9}
\Psi = S +\lambda \sqrt{M} ,
\end{eqnarray}
\begin{eqnarray}
\lambda = \pm 1 ,
\end{eqnarray}
\begin{eqnarray}\label{5.4.10}
a = \frac{12p^{2}( \cos^{2}\delta + D \sin^{2}\delta)}{B'},
\end{eqnarray}
\begin{eqnarray}\label{5.4.11}
M = 1 + \frac{12p^{2}B''( \cos^{2}\delta
+D \sin^{2}\delta)}{B'^{2}},
\end{eqnarray}
\begin{eqnarray}\label{5.4.12}
Z = \xi \sin\delta + \zeta \cos\delta - U \tau.
\end{eqnarray}
For the existence of the solitary wave solution (\ref{5.4.6}), it is necessary that the following condition is satisfied.
\begin{eqnarray}\label{5.4.7}
L = MB'^{2} = B'^{2} + 12B''p^{2}( \cos^{2}\delta + D
\sin^{2}\delta)>0.
\end{eqnarray}
If the condition (\ref{5.4.7}) holds good, $U$ is given by the
equation
\begin{eqnarray}\label{5.4.13}
U=4p^{2}a_{3},
\end{eqnarray}
where
\begin{eqnarray}\label{5.4.3}
a_{3} &=& \frac{1}{2}A\cos\delta(\cos^{2}\delta+D\sin^{2}\delta).
\end{eqnarray}
With the help of the equations (\ref{5.4.10}), (\ref{5.4.11}), (\ref{5.4.13}) and (\ref{5.4.3}), we get the following expressions of $p$, $U$ and $M$ to express them in terms of $a$.
\begin{eqnarray}\label{5.43}
p &=& \frac{1}{2}\sqrt{\frac{aa_{1}}{6a_{3}}} , \\ U &=& \frac{1}{6}aa_{1},\\
M &=& 1+a \frac{a_{2}}{a_{1}},
\end{eqnarray}
where
\begin{eqnarray}\label{54a.3}
a_{1} &=& A B' cos\delta , \\ a_{2} &=& AB''\cos\delta.
\end{eqnarray}
Using (\ref{5.43}), we can write the Eq.(\ref{5.4.6}) as
\begin{eqnarray}\label{5.4.14}
\varphi ^{(1)} &=&\varphi _{0}(Z)\nonumber\\
&=&a\frac{sech\bigg[\sqrt{\frac{aa_{1}}{6a_{3}}}(\xi \sin\delta + \zeta
\cos\delta -
\frac{1}{6}aa_{1}\tau)\bigg]}{sech\bigg[\sqrt{\frac{aa_{1}}{6a_{3}}}(\xi
\sin\delta + \zeta \cos\delta -
\frac{1}{6}aa_{1}\tau)\bigg]+\lambda\sqrt{1+a \frac{a_{2}}{a_{1}}}}.
\end{eqnarray}
Assuming that $a$ to be a slowly varying function of time, following Ott and Sudan
\cite{Ott69}, we introduced the following space coordinate in a frame moving
with the solitary wave.
\begin{eqnarray}\label{5.4.23}
\overline{Z}=\sqrt{\frac{aa_{1}}{6a_{3}}}\Bigg(\xi sin\delta + \zeta cos\delta
- \frac{1}{6}a_{1}\int^{\tau}_{0}ad\tau\Bigg).
\end{eqnarray}
It is important to note that if $a$ is a constant, then $\overline{Z}=2pZ$ and consequently,
\begin{eqnarray}\label{5.414}
\varphi ^{(1)}&=&\varphi _{0}(\overline{Z})\nonumber \\
&=& a \frac{sech\overline{Z}}{sech\overline{Z}+\lambda\sqrt{M}}\nonumber\\
&=&a\frac{sech\bigg[\sqrt{\frac{aa_{1}}{6a_{3}}}(\xi \sin\delta + \zeta
\cos\delta - \frac{1}{6}a_{1}\int^{\tau}_{0}ad\tau)\bigg]}{sech\bigg[\sqrt{\frac{aa_{1}}{6a_{3}}}(\xi
\sin\delta + \zeta \cos\delta - \frac{1}{6}a_{1}\int^{\tau}_{0}ad\tau)\bigg]+\lambda\sqrt{1+a \frac{a_{2}}{a_{1}}}}.
\end{eqnarray}
is the solitary wave solution of the combined MKdV-KdV-ZK equation propagating at an angle $\delta$ to the external uniform static magnetic field. Now dropping ``overline'' on $\overline{Z}$, we can write the equation (\ref{5.414}) as
\begin{eqnarray}\label{5.415}
\varphi ^{(1)} &=& \varphi _{0}(Z)\nonumber\\
&=& a \frac{sechZ}{sechZ+\lambda\sqrt{M}}\nonumber\\
&=&a\frac{sech\bigg[\sqrt{\frac{aa_{1}}{6a_{3}}}(\xi\sin\delta+\zeta
\cos\delta-
\frac{1}{6}a_{1}\int^{\tau}_{0}ad\tau)\bigg]}{sech\bigg[\sqrt{\frac{aa_{1}}{6a_{3}}}(\xi
\sin\delta+\zeta\cos\delta-
\frac{1}{6}a_{1}\int^{\tau}_{0}ad\tau)\bigg]+\lambda\sqrt{1+a \frac{a_{2}}{a_{1}}}},
\end{eqnarray}
where $Z$ is given by the following equation:
\begin{eqnarray}\label{5.423}
Z=\sqrt{\frac{aa_{1}}{6a_{3}}}\Bigg(\xi sin\delta + \zeta cos\delta
- \frac{1}{6}a_{1}\int^{\tau}_{0}ad\tau\Bigg).
\end{eqnarray}
Now our aim is to find the condition for which $\varphi^{(1)}$ given by the equation (\ref{5.415}) is a solitary wave solution of the further modified macroscopic equation (\ref{5.3.19}).

With the change of variable defined by the equation (\ref{5.423}) and assuming that $\varphi^{(1)}$ is a function of $Z, \tau$ only, Eq.(\ref{5.3.19}) can be written as
\begin{eqnarray}\label{5.4.24}
&&\frac{\partial
\varphi^{(1)}}{\partial\tau}+\bigg(-\frac{1}{3}a_{1}pa+\frac{Z}{2a}\frac{\partial
a}{\partial\tau}\bigg)\frac{\partial\varphi^{(1)}}{\partial
Z}+2pa_{1}\varphi^{(1)}\frac{\partial\varphi^{(1)}}{\partial
Z}+2pa_{2}(\varphi^{(1)})^{2}\frac{\partial\varphi^{(1)}}{\partial
Z}\nonumber\\&&+8p^{3}a_{3}\frac{\partial^{3}\varphi^{(1)}}{\partial
Z^{3}}+AE\alpha_{2}p\cos\delta
P\int^{\infty}_{-\infty}\frac{\partial\varphi^{(1)}}{\partial
Z}\frac{\partial Z'}{Z-Z'}=0.
\end{eqnarray}
To investigate the solution of Eq. (\ref{5.4.24}), we follow Ott and
Sudan \cite{Ott69} and generalizing the multiple-time scale analysis with
respect to $\alpha_{2}$, by setting
\begin{eqnarray}\label{5.4.26}
\varphi^{(1)}(Z,\tau)=q^{(0)}+\alpha_{2}q^{(1)}+\alpha_{2}^{2}q^{(2)}+\alpha_{2}^{3}q^{(3)}+.........
\end{eqnarray}
where each $q^{(j)} (j=0,1,2,3,....)$ are the function of
$\tau=\tau_{0},\tau_{1},\tau_{2}.....$ . Here $\tau_{j}$ is given by
\begin{eqnarray}\label{5.4.27}
\tau_{j}=\alpha_{2}^{j}\tau, j=0,1,2,3,........
\end{eqnarray}
Substituting (\ref{5.4.26}) into (\ref{5.4.24}) and then equating
the coefficient of different power of $\alpha_{2}$ on each side of
Eq. (\ref{5.4.24}), we get a sequence of equations. The zeroth
and the first order equation of this sequence are respectively, given by the following equations.
\begin{eqnarray}\label{5.4.28}
\rho\bigg[\frac{\partial}{\partial\tau}+\frac{Z}{2a}\frac{\partial
a}{\partial\tau}\frac{\partial}{\partial
Z}\bigg]q^{(0)}+\overline{L}\frac{\partial }{\partial Z}q^{(0)}=0,
\end{eqnarray}
\begin{eqnarray}\label{5.4.29}
\rho\bigg[\frac{\partial}{\partial\tau}+\frac{Z}{2a}\frac{\partial
a}{\partial\tau}\frac{\partial}{\partial
Z}\bigg]q^{(1)}+\frac{\partial }{\partial
Z}\overline{L}q^{(1)}=\rho\overline{M}q^{(0)},~
\end{eqnarray}
where
\begin{eqnarray}\label{5.4.30}
\overline{L}=\frac{\partial^{2}}{\partial
Z^{2}}+\frac{6}{a}~q^{(0)}+\frac{6(M-1)}{a^{2}}[q^{(0)}]^{2}-1,~
\end{eqnarray}
\begin{eqnarray}\label{5.4.31}
\rho=6\sqrt{\frac{6a_{3}}{a_{1}^{3}}}~a^{-\frac{3}{2}},
\end{eqnarray}
\begin{eqnarray}\label{5.4.32}
\overline{M}q^{(0)}=&-&\bigg[\frac{\partial q^{(0)}}{\partial
\tau_{1}}+\frac{Z}{2a}\frac{\partial a}{\partial
\tau_{1}}\frac{\partial q^{(0)}}{\partial Z}+AEp\cos\delta
P\int^{\infty}_{-\infty}\frac{\partial q^{(0)}}{\partial
Z'}\frac{\partial Z'}{Z-Z'}\bigg].
\end{eqnarray}
Now it can be easily verified that
$q^{(0)}=a\frac{sechZ}{sechZ+\lambda\sqrt{M}}$ is the soliton
solution of the zeroth order equation if
\begin{eqnarray}\label{5.4.33}
\frac{\partial a}{\partial \tau}=0 ,
\end{eqnarray}
which implies that $a$ is independent of time, i.e., at the lowest order, the solitary wave solution of the further modified macroscopic evolution equation is same as that of the combined MKdV-KdV-ZK equation.

\begin{figure}
  \begin{center}
  \includegraphics{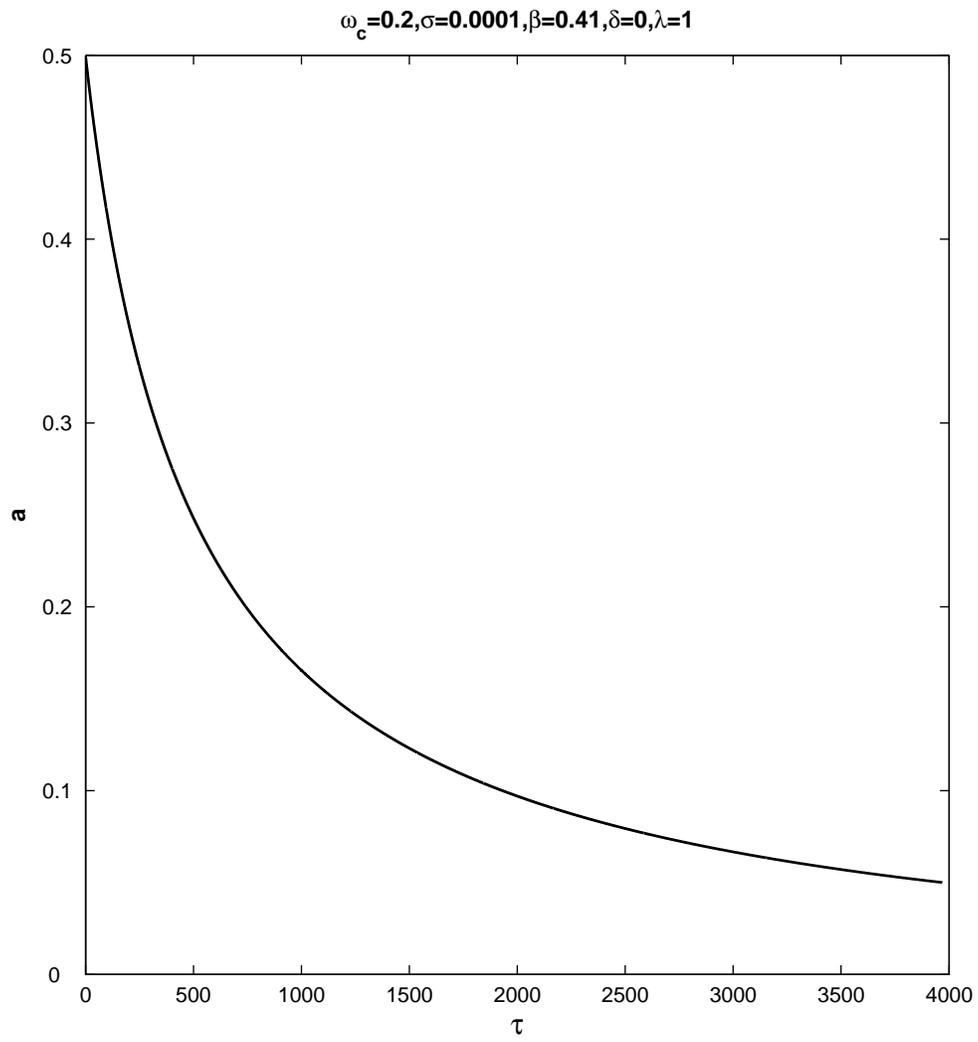}
  \caption{Variation of $a$ against $\tau$}\label{fig:new}
  \end{center}
\end{figure}

Using (\ref{5.4.33}), Eq.(\ref{5.4.29}) can be written as
\begin{eqnarray}\label{5.4.34}
\rho\frac{\partial q^{(1)}}{\partial\tau}+\frac{\partial }{\partial
Z}\overline{L}q^{(1)}=\rho\overline{M}q^{(0)}.
\end{eqnarray}
Now for the existence of a solution of the equation (\ref{5.4.34}), its right
hand must be perpendicular to the kernel of the operator adjoint to
the operator $\frac{\partial }{\partial Z}\overline{L}$; this kernel, which
must tend to zero as $|Z|\rightarrow \infty$ is
$\frac{sechZ}{sechZ+\lambda\sqrt{M}}$. Thus we get the following consistency condition for the existence of a solution of the equation (\ref{5.4.34}).
\begin{eqnarray}\label{5.4.35}
\int^{\infty}_{-\infty}\frac{sechZ}{sechZ+\lambda\sqrt{M}}~\overline{M}q^{(0)}dZ=0.
\end{eqnarray}
From equation (\ref{5.4.35}), we get the following differential equation
for the solitary wave amplitude $a$.
\begin{eqnarray}\label{5.4.36}
&&\frac{\partial a}{\partial
\tau_{1}}+\frac{AEa^{3/2}(B'+aB'')\cos\delta}{\sqrt{3B'(\cos^{2}\delta+D
\sin^{2}\delta) }}
P\int_{-\infty}^{\infty}\int_{-\infty}^{\infty}\Psi(Z)\frac{\partial}{\partial
Z'}[\Psi(Z')]\frac{dZ'dZ}{Z-Z'}=0,
\end{eqnarray}
where
\begin{eqnarray}
\Psi(Z)=\frac{sechZ}{sechZ+\lambda\sqrt{M}}.
\end{eqnarray}
Using the relation $\tau_{1}=\alpha_{2}\tau$, the equation (\ref{5.4.36}) can be written in  the following simplified form:
\begin{eqnarray}\label{5.436}
&&\frac{\partial a}{\partial
\tau}+\frac{AE \alpha_{2}a^{3/2}(B'+aB'')\cos\delta}{\sqrt{3B'(\cos^{2}\delta+D
\sin^{2}\delta) }}
P\int_{-\infty}^{\infty}\int_{-\infty}^{\infty}\Psi(Z)\frac{\partial [\Psi(Z')]}{\partial
Z'}\frac{dZ'dZ}{Z-Z'}=0.
\end{eqnarray}
Here it is important to note that $M(=1+a \frac{a_{2}}{a_{1}})$ appearing in $\Psi(Z)$ is a function of $a$. So, it is not possible to find the exact analytical dependence of $a$ on $\tau$. But we can solve the above equation by using the Taylor series expansion for the terms of the form $\frac{1}{sechx+\lambda \sqrt{M}}$ in powers of $a$. Keeping terms upto the order $a^{5/2}$, we get the following differential equation for $a$ from equation (\ref{5.436}).
\begin{eqnarray}\label{5.4.37}
\frac{\partial a}{\partial
\tau} &+& AE \alpha_{2}a^{\frac{3}{2}}\sqrt{\frac{B'}{3(\cos^{2}\delta+D
\sin^{2}\delta) }}\cos\delta~
\gamma_{1}\nonumber \\
&-& AE\alpha_{2}a^{\frac{5}{2}}\frac{\lambda}{2}\frac{B''}{\sqrt{3B'(\cos^{2}\delta+D
\sin^{2}\delta)}}\cos\delta~
(\gamma_{2}+\gamma_{3})\nonumber \\
&+& AE\alpha_{2}a^{\frac{5}{2}}\frac{B''}{\sqrt{3B'(\cos^{2}\delta+D
\sin^{2}\delta)}}\cos\delta~
\gamma_{1}= 0,
\end{eqnarray}
where $\gamma_{1}$, $\gamma_{2}$, $\gamma_{3}$ are given by the following integrals.
\begin{eqnarray}
\left.\begin{array}{ll} 
\gamma_{1} = P\int^{\infty}_{-\infty}\int^{\infty}_{-\infty}\Phi_{1}(Z)\frac{\partial}{\partial
Z'}[\Phi_{1}(Z')]\frac{dZ'dZ}{Z-Z'},\\\\
\gamma_{2} = P\int^{\infty}_{-\infty}\int^{\infty}_{-\infty}\Phi_{1}(Z)\frac{\partial}{\partial
Z'}[\Phi_{2}(Z')]\frac{dZ'dZ}{Z-Z'},\\\\
\gamma_{3} = P\int^{\infty}_{-\infty}\int^{\infty}_{-\infty}\Phi_{2}(Z)\frac{\partial}{\partial
Z'}[\Phi_{1}(Z')]\frac{dZ'dZ}{Z-Z'}.
\end{array}\right\}
\end{eqnarray}
$\Phi_{1}(Z)$ and $\Phi_{2}(Z)$ appearing in the above are given by
\begin{eqnarray}
\Phi_{1}(Z) &=& \frac{sechZ}{sechZ+\lambda},\\
\Phi_{2}(Z) &=& \frac{sechZ}{(sechZ+\lambda)^{2}}.
\end{eqnarray}
Now solving the above differential equation (\ref{5.4.37}) for $a$ by the use of the initial condition, $a=a_{0}$ when $\tau = 0$, we get the following equation for $a$:
\begin{eqnarray}\label{5.4.38}
\mu \tan^{-1}\Bigg[\frac{\mu
(\sqrt{a}-\sqrt{a_{0}})}{1+\mu^{2}\sqrt{aa_{0}}}\Bigg]
-\frac{\sqrt{a}-\sqrt{a_{0}}}{\sqrt{aa_{0}}}=\frac{\tau}{\tau'},\nonumber\\
\end{eqnarray}
where
\begin{eqnarray}\label{5.4.40}
\tau'=\bigg[\frac{1}{2}AE \alpha_{2} \sqrt{\frac{B'}{3(\cos^{2}\delta+D
\sin^{2}\delta) }}\cos\delta~ \gamma_{1}\bigg]^{-1},
\end{eqnarray}
\begin{eqnarray}\label{5.4.39}
\mu=\sqrt{\frac{B''}{B'}\bigg[1-\frac{\lambda}{2}\frac{(\gamma_{2}+\gamma_{3})}{\gamma_{1}}\bigg]}.
\end{eqnarray}
From equation (\ref{5.4.38}), we see that $a$ is implicitly depends on $\tau$ and consequently, from this equation it is not possible to predict the nature (decreasing or increasing) of dependence of $a$ on $\tau$. But plotting $a$ against $\tau$ for the appropriate set of values of the parameters involved in the system, we find that $a$ is slowly varying function of time. By the phrase `` appropriate set of values of the parameters'', we mean that those values of the parameters of the system for which the condition for existence of alternative solitary wave solution of the combined MKdV-KdV-ZK equation holds good, i.e., for those values of the parameters of the system for which $L>0$. Taking $a_{0}=0.5$ (arbitrary) and the values of the parameters as mentioned in the figure \ref{fig:new}, we plot $a$ against $\tau$ in Fig.\ref{fig:new}. This figure clearly shows that the amplitude ($a$) decays slowly with time ($\tau$) and consequently, the amplitude of the alternative solitary wave solution of the combined MKdV-KdV-ZK equation is a slowly varying function of time when the effect of Landau damping is considered.

\section{\label{conclusion}Conclusions}

A macroscopic evolution equation corresponding to the combined MKdV-KdV-ZK equation has been derived to include the effect of Landau damping. This macroscopic evolution equation admits the same alternative solitary wave solution of the combined MKdV-KdV-ZK equation except the fact that the amplitude of the solitary wave solution of the combined MKdV-KdV-ZK like macroscopic equation is a slowly varying function of time. The multiple time scale method of Ott and Sudan \cite{Ott69} has been generalized here to solve the said evolution
equation. In small amplitude limit, we have observed the following result.
\begin{description}
    \item[Result:] Due to inclusion of the effect of Landau damping, the amplitude of the alternative solitary
    wave solution having profile different from $sech^{2}$ or $sech$ of the macroscopic evolution equation decays
    slowly with time.
\end{description}

\bibliography{aps_jdabkpd}

\newpage

\newpage

%

\end{document}